\documentstyle[NATO,epsf,namedreferences]{CRCKAPB} 
 
\begin{opening}
\title{HIDDEN sd/wd STARS AMONG THE FAUST UV SOURCES TOWARD OPHIUCHUS}
\author{LILIANA FORMIGGINI}
\institute{\it School of Physics and Astronomy and The Wise Observatory\\
Tel-Aviv University, Israel}
\end{opening}

\begin{document}

No white dwarf or hot subdwarf stars are found as optical counterparts
of the 228 UV sources detected in a UV image toward Ophiuchus (Formiggini et al. 2002).
The image was  obtained at 1600 \AA , with the FAUST instrument 
on board of the space shuttle  Atlantis. 
Hot subluminous stars are numerous among blue objects, and dominate the population 
of blue stars down to B=16.5. 
To search for such stars,
we cross-identified the UV sources positions with objects in the Main Part 
of the Tycho catalog, finding a significant parallax value for 46 entries.
A significant number of objects have  M${_V}$ values as 
expected for WD or subdwarf  stars (Fig. 1a), while their spectral classification
spans types from B to early F.
For all these stars the parallax errors are very large, hence their 
classification as  subluminous stars on the basis of the  parallax data 
is rather unreliable.

\begin{figure}[h]
\epsfbox{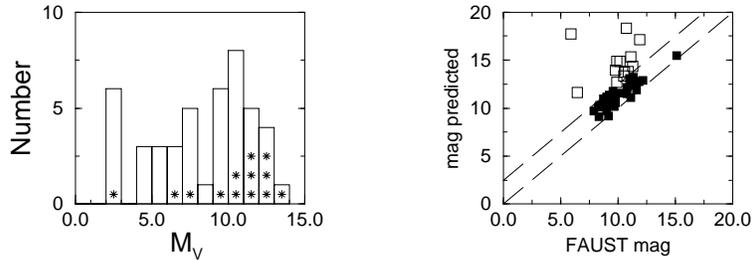}
\caption{\bf{a) Hystogram of the calculated M${_V}$ magnitudes; b) FAUST versus predicted magnitudes}}
\end{figure}

Synthetic photometry of spectral data was performed in order to predict the
expected UV emission (Fig. 1b). Although the predicted magnitudes are
systematically fainter than the measured ones, the majority of the stars
populate a region inside the dotted lines. 
The open squares show the stars for which FAUST measured a brighter UV 
 magnitude than the one predicted on the basis of the photometric color. 
These uv-excess (also marked by asterisk in Fig. 1a)  should belong to the subdwarf 
or white dwarf luminosity classes but one, already classified as an Am star.
Nine of the twelve uv-excess stars with M${_V}$ $\ge$ 5 were observed at 
the 1.0 meter telescope at the Wise Observatory. For all, but one, the Balmer
lines are moderately broadened.
The dashed line in Fig 2 is  the model profile from Wesemael et al. (1980). 
Two of the observed candidates can be fitted with atmospheres of log g $\ge 7 $, 
as for white dwarfs stars, and six are sdB stars.

\begin{figure}[h]
\epsfbox{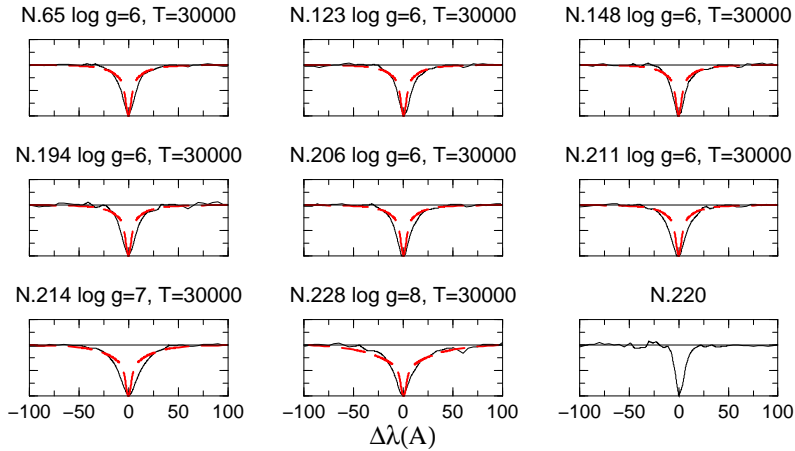}
\caption{ \bf H$\beta$ line profile observed (line) compared with
model atmospheres profile (dashed). }
\end{figure}

It is known that the PG survey is quite incomplete at
bright magnitudes. This analysis  shows that combining the UV information with even  
poorly determined parallaxes, some new sd/wd stars are detected and that indeed more such 
bright stars are still unrecognized in the existing samples.
Extending this analysis to the other FAUST fields previously identified at 
the Wise Observatory, more bright sd/wd stars, 
suitable for a determination of the space density at the bright end of the 
luminosity function can be identified.


\begin{thebibliography}{}
\bibitem [] {} 
Formiggini, L., Brosch, N., Almoznino, E., Bowyer, S. and Lampton, M., 2002, 
 {\it MNRAS\/}, {\bf 332}, 441.

\bibitem [] {}
Wesemael F., Auer L.H., Van Horn H.M. and Savedoff M.P.  1980, {\it ApJS\/}, {\bf 43},159.
\end{thebibliography}
\end{document}